      [1999/12/01 v1.4c Il Nuovo Cimento]
\documentclass{cimento}

\usepackage{graphicx}
\usepackage{amsmath} 
\title{$CP$ violation in top-quark physics}
\author{German~Valencia\from{ins:x} }
\instlist{\inst{ins:x} Department of Physics, Iowa State University, Ames, IA 50011  }
\PACSes{\PACSit{12.15.Ji, 12.15.Mm, 12.60.Cn, 13.20.Eb,
13.20.He, 14.70.Pw}{}}

\begin{document}

\maketitle

\begin{abstract}
In this talk I discuss how to search for CP violation in top-quark pair production and decay using T-odd correlations. I discuss two examples  which illustrate many of the relevant features: CP violation in a heavy neutral Higgs boson; and CP violating anomalous top-quark couplings. I present some numerical results for the LHC and some for the Tevatron.

\end{abstract}

\section{Introduction}

CP violation beyond the standard model (SM) has yet to be observed but we suspect it must be there in order to explain the baryon asymmetry of the universe. This gives paramount importance to new searches for CP violation in the high energy frontier. One tool, proposed many years ago, for searches in collider experiments is the use of triple product correlations \cite{tprods}. These are simple kinematic correlations of the form $\vec{p}_1\cdot(\vec{p}_2\times\vec{p}_3)$. 

These correlations are referred to as ``naive-T'' odd because they  reverse sign under the ``naive-T'' operation that reverses the direction of momenta and spin without interchanging initial and final states. These correlations do not have to be CP-odd, they can be induced by CP conserving interactions because the naive-T operation does not correspond to the time reversal operation. It is well known, however, that CP conserving T-odd correlations only occur {\it beyond tree level}, and for this reason we will refer to them as being induced by ``unitarity phases''. This means that the CP conserving background is both small and interesting in its own right. The CP nature of a given correlation can be determined easily as we will show with examples later on. An important point is that the generic momenta $\vec{p}_i$ that enters the correlation can be that of a composite object, such as a jet.

Triple product correlations appear in the calculation of invariant matrix elements as contractions of four independent four-vectors with the Levi-Civita tensor. This immediately indicates that unless one is discussing effective theories containing vertices that couple more than four particles, the correlations will always appear first as spin correlations involving intermediate states.  Weak decays of intermediate particles  will then act  as spin analyzers yielding triple product correlations that only involve momenta. From the simple properties of triple product correlations we see that the top-quark pair production and decay processes  in colliders are an ideal laboratory to investigate these observables \cite{ttpairs}.

\begin{figure}
\hspace{0.7in}\includegraphics[width=1.5in,angle=90]{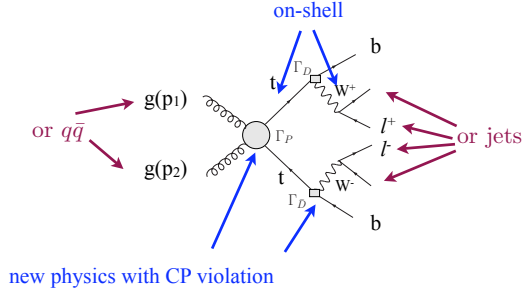}   
\caption{Kinematic configuration: the top-quark and $W$-bosons are treated on-shell and CP violation due to new physics is allowed in both production and decay processes for top-quarks only.}
\label{fig1}
\end{figure}

A generic diagram for the processes we consider is shown in Figure~\ref{fig1}. The circle in the production process represents the top-quark pair production including the SM diagrams and CP violation due to new physics. Similarly the square in the top decay process represents top decay via the SM and the additional CP violating interactions. The $W$ decay is assumed to proceed as in the SM and we will consider  both the leptonic and hadronic (jets) cases.

\begin{figure}
\hspace{0.1in}\includegraphics[width=1.5in,angle=270]{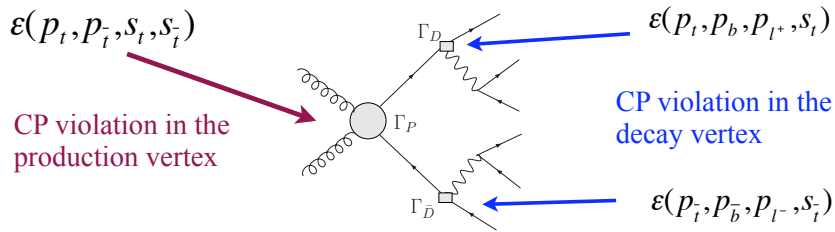}   
\caption{Kinematic configuration: the T odd correlations originate as spin correlations in the production and/or decay vertices.}
\label{fig2}
\end{figure}

In Fig.~\ref{fig2} we see how the topology of the processes we discuss implies that the T-odd correlations originate as spin correlations. When CP violation occurs in the production vertex, the only Lorentz invariant (scalar) T-odd correlation that occurs is one that involves the momenta and spin of both the top and anti-top. The weak decays then act as spin analyzers and the top-spin gets replaced by one of the final state momenta in the final correlation. In the case of CP violation in the decay vertex we see that the induced spin correlations are not CP eigenstates. A comparison of both decay vertices is required to separate CP violation from signals induced by unitarity phases.

\section{Sources of CP violation}

To illustrate the different possibilities we will consider two simple examples of CP violation. The first one will be a heavy (non-SM) Higgs boson with simultaneous scalar and pseudo-scalar couplings to the top-quark. This example will illustrate the possibility of large intrinsic asymmetries. The second example will be the CP violating anomalous top-quark couplings. This example will generate more general T-odd correlations, involving momenta from the production and decay processes. It will provide examples of both CP-odd and CP-even observables. The disadvantage of this case is that the asymmetries are small by assumption, because the anomalous couplings are necessarily small (to remain a valid description of the top-quark couplings).

\subsection{CP violation via neutral Higgs}

A neutral Higgs has a generic coupling to top-quarks given by ${\cal L}=-m_t/v H\bar{t}(A+iB\gamma_5)t$, which violates CP if both $A$ and $B$ are non-zero simultaneously. This kind of coupling occurs in multi-Higgs models where Weinberg has shown that unitarity requires $|AB|\leq 1/\sqrt{2}$ \cite{Weinberg:1990me}. To reach this upper bound requires a few conditions on the models:  the lightest neutral mass eigenstate must be dominant and all vevs must have similar sizes. Here we will not concern ourselves with these details and simply   assume that this bound is saturated for our numerical studies (as is also done in many previous papers \cite{other}.)

We will consider the decay chain $H\to t\bar{t} \to b\bar{b} W^+ W^-$ assuming for the time being that the $W$ bosons decay hadronically and are reconstructed as one jet. Under these conditions there is only one T-odd correlation and it is given by
\begin{eqnarray}
\def\cpto{\mathrel {\vcenter {\baselineskip 0pt \kern 0pt
    \hbox{$H_{r.f.}$} \kern 0pt \hbox{$\longrightarrow$} }}}
{\cal O}_1\ =\ \epsilon(p_t,p_{\bar {t}},p_b,p_{\bar {b}}) \,\,
\xrightarrow[]{H_{C.M.}}\,\, \propto  \vec{p}_t\
\cdot  \left(\vec{p}_b\times\vec{p}_{\bar b}\right) \nonumber \\
\xrightarrow[]{CP}\,\, -\vec{p}_{\bar t}\
\cdot  \left(-\vec{p}_{\bar b}\times-\vec{p}_b\right) \,\, = \,\,
-\vec{p}_t\
\cdot  \left(\vec{p}_b\times\vec{p}_{\bar b}\right).
\end{eqnarray}
The arrow in the first line shows the form taken by the Lorentz scalar in the Higgs rest frame: the triple product correlation form is frame specific but arises from a Lorentz scalar that can be evaluated in any frame. The second line sketches the proof (in the Higgs rest frame) that this correlation is indeed CP odd.
To measure the correlation we use a counting asymmetry such as
\begin{eqnarray}
A_{CP}=\frac{N_{events}(\vec{p}_t\
\cdot  \left(\vec{p}_b\times\vec{p}_{\bar b}\right)>0)-N_{events}(\vec{p}_t\
\cdot  \left(\vec{p}_b\times\vec{p}_{\bar b}\right)<0)}{N_{events}(\vec{p}_t\
\cdot  \left(\vec{p}_b\times\vec{p}_{\bar b}\right)>0)+N_{events}(\vec{p}_t\
\cdot  \left(\vec{p}_b\times\vec{p}_{\bar b}\right)<0)}
\end{eqnarray}
or we can directly fit the differential decay distribution to extract terms linear in ${\cal O}_1$. The case of Higgs decay is very simple, and a fully analytic result is possible \cite{Valencia:2005cx}.  The maximum value (when Weinberg's unitarity bound is saturated) $A_{CP}\sim 7\%$  indicates that large intrinsic asymmetries are possible. 

When we go beyond Higgs decay and consider top pair production at LHC the signal is diluted by the larger source of top-quark pairs (gluon fusion without a Higgs) without CP violation. This exercise does show why the CP test worked even though the initial $pp$ state at LHC is not a CP eigenstate. The LHC simply served as a Higgs factory and the CP properties were fixed by the final state alone. The argument extends to the case of top-quark pair production from gluon fusion of $q\bar{q}$ annihilation, and fails for $qq$ initial states (that can produce $t\bar{t}$ pairs plus other particles). The later, however, has been estimated to be a very small background \cite{review}. 

\subsection{Top-quark anomalous couplings}

We now discuss the more general example of top-quark anomalous couplings. CP violating couplings of this sort are, for example, the top-quark electric dipole moment (EDM) or color electric dipole moment (CEDM). These couplings are negligibly small in the SM but can be much larger in general. In particular, models in which CP violation is induced by the exchange of neutral scalars have contributions to the EDM (or CEDM) of fermions that scale as $m_f^3$, suggesting a potentially large top-quark CEDM. To consider the CEDM of the top-quark we start from the effective Lagrangian
\begin{equation}
{\cal L}_{cedm}=-ig_s\frac{\tilde{d}}{2}\ \bar{t} \ \sigma_{\mu\nu}\gamma_5\ t\ G^{\mu\nu}
\label{topcedm}
\end{equation}
which modifies the $ttg$ vertex but also introduces a ``seagull'' $ttgg$ term that is required by gauge invariance. The contributions of these two vertices to top-quark production have been written down in a Lorentz invariant form \cite{Antipin:2008zx}. For example, for the leptonic decay of both $t$ and $\bar{t}$ they result in three different T-odd and CP-odd correlations
\begin{eqnarray}
{\cal O}_1 &=& \epsilon(p_t,p_{\bar{t}},p_{\mu^+},p_{\mu^-}) \nonumber \\
{\cal O}_2 &=& \,(t-u) \,\epsilon(p_{\mu^+},p_{\mu^-},P,q) \nonumber \\
{\cal O}_3 &=&\,(t-u) \,\left( P \cdot p_{\mu^+} \,
\epsilon(p_{\mu^-},p_t,p_{\bar{t}},q)+P \cdot p_{\mu^-} 
\,\epsilon(p_{\mu^+},p_t,p_{\bar{t}},q) \right),
\end{eqnarray}
where the sum and difference of parton momenta are denoted by $P$ and $q$ respectively. These formula can be easily adapted to semileptonic and purely hadronic channels. For example for $W$'s reconstructed as one jet the lepton momentum is replaced by the $b$-jet; for hadronic $W$ decay it is replaced by the $d$-jet momentum, etc \cite{Gupta:2009wu,Gupta:2009eq}. Notice that the correlations are quadratic in $q$ as they need to be for two  indistinguishable initial state particles. The CP-odd part of the differential cross-section is expressed in terms of these three correlations as
\begin{equation}
d\sigma \sim C_1(s,t,u){\cal O}_1 + C_2(s,t,u){\cal O}_2 +C_3(s,t,u){\cal O}_3
\end{equation}
where the form factors $C_i$ can be found in Ref.~\cite{Antipin:2008zx}. This expression with three independent CP-odd and T-odd correlations appears to be the most general one, although a formal proof is not available. 

Additional $tbW$ anomalous couplings can introduce CP violation in the decay vertex, we write it as
\begin{eqnarray}
\Gamma^\mu_{Wtb} &=& 
-\frac{g}{\sqrt{2}} \, V_{tb}^\star \,\bar{u}(p_b) \left[ \gamma_\mu (f_1^L P_L+f_1^R P_R)-
i  \sigma^{\mu\nu} (p_t-p_b)_\nu (f_2^L P_L+f_2^R P_R)\right] u(p_t). \label{ftilde}
\end{eqnarray}
This vertex can be derived from a dimension five effective Lagrangian as in Ref.~\cite{delAguila:2002nf}, but unlike the case of Eq.~\ref{topcedm}, the effective Lagrangian does not generate other vertices that affect this calculation. Numerically we use $V_{tb}\equiv 1$, $f_1^L=1$, $f_1^R=0$ and $f_2^L=0$ as in the SM, and allow for new physics only through the coupling $f_2^R$ which is the only one that can interfere with the SM to produce $T$-odd correlations in the limit when $m_b=0$.To generate  $T$-odd observables the coupling $f_2^R$ must have a phase but this phase does not have to be $CP$ violating. We thus write $f_2^R=f\exp{i(\phi_f+\delta_f)}$ using $\phi_f$ to parametrize a $CP$ violating phase due to new physics and $\delta_f$ a $CP$ conserving phase arising from real intermediate states at the loop level. 

The spin and color averaged matrix element squared containing the $T$-odd correlations in this case looks like~\cite{Antipin:2008zx} \begin{eqnarray}
|{\cal M}|^2_{T} &=& \, 
f\sin(\phi_f+\delta_f)\, \epsilon(p_t,p_{ b},p_{\ell^+},Q_{t}) +
 f\sin(\phi_f-\delta_f) \,\epsilon(p_{\bar t},p_{ \bar{b}},p_{\ell^-},Q_{\bar{t}}) .
 \label{asymcpdec}
\end{eqnarray} 
All the terms in Eq.~\ref{asymcpdec} contain three four-momenta from one of the decay vertices and a fourth ($Q$) `spin-analyzer' which is a linear combination of other momenta in the reaction (its precise form can be found in Ref.~\cite{Antipin:2008zx}). Note that these correlations are not CP-odd as was the case for CP violation in the production vertex. One needs to compare the top and anti-top decays to extract either a CP-odd observable or a CP-even one.

\section{Observables}

After obtaining the `theoretical' correlations, we must find observable correlations. We define these ones as those that involve only the following momenta: lepton ($p_{\mu^\pm}$); $b$-jet ($p_{b,\bar{b}}$); beam momentum ($\tilde{q}\equiv P_1-P_2$); non-$b$ jet momenta ordered by $p_T$ ($p_{j1},p_{j2}\cdots$). Any other CP blind ordering of the non-$b$ jets will also work. A few of the correlations discussed in Refs.~\cite{Gupta:2009wu,Gupta:2009eq} for different cases are, respectively:

\begin{itemize}

\item Dimuon events at LHC: CP-odd correlations
\begin{eqnarray}
\tilde {\cal{O}}_1 &=& \epsilon(p_b,p_{\bar{b}},p_{\mu^+},p_{\mu^-}) \,\,
\xrightarrow[]{b\bar b ~CM}\,\, \propto \,\, \vec{p}_b\cdot (\vec{p}_{\mu^+} \times \vec{p}_{\mu^-})\nonumber \\
\tilde {\cal{O}}_2 &=& \, \tilde{q}\cdot (p_{\mu^+}-p_{\mu^-}) \,\epsilon(p_{\mu^+},p_{\mu^-},p_b+p_{\bar{b}},\tilde{q}) 
\end{eqnarray}
The second from for the first correlation specializes to the $b\bar{b}$ center of mass frame where the correlation takes the form of a simple triple product and where the CP properties are evident. The second correlation provides an example in which it is not necessary to distinguish the $b$ and $\bar{b}$ jets; it is also quadratic in $\tilde{q}$ as needed for identical particles in the initial state. A detailed study of a correlation proportional to $\tilde{\cal O}_1$ for Atlas has been performed by J. Sj\"{o}lin \cite{Sjolin:2003ah}.

\item Dimuon events at LHC: CP-even T-odd correlation to study absorptive phases without the need to distinguish $b$ from $\bar{b}$:
\begin{eqnarray}
{\cal O}_a &=& \, \tilde{q}\cdot (p_{\mu^+}+p_{\mu^-}) \,\epsilon(p_{\mu^+},p_{\mu^-},p_b+p_{\bar{b}},\tilde{q}).
\end{eqnarray}

\item Muon plus jets events at the Tevatron: CP-odd correlations. Notice some require distinguishing $b$ from $\bar{b}$ but some don't.
\begin{eqnarray}
{\cal O}_2 &=& \epsilon(P,p_b+p_{\bar b},p_\ell,p_{j1}) \,\,
\xrightarrow[]{lab}\,\, \propto \,\, (\vec{p}_b +\vec{p}_{\bar b})\cdot (\vec{p}_\ell \times \vec{p}_{j1})
\nonumber \\
{\cal O}_3 &=&  Q_\ell \, \epsilon(p_b,p_{\bar b},p_\ell,p_{j1})  \,\,
\xrightarrow[]{b\bar b ~CM}\,\, \propto \,\, Q_\ell\,\vec{p}_b \cdot (\vec{p}_\ell \times \vec{p}_{j1})
\nonumber \\
{\cal O}_7 &=&  \tilde{q}\cdot(p_b-p_{\bar b})\, \epsilon(P,\tilde{q},p_b,p_{\bar b}) \,\,
\xrightarrow[]{lab}\,\, \propto \,\, \vec{p}_{beam}\cdot (\vec{p}_b -\vec{p}_{\bar b})  \,\vec{p}_{beam}\cdot (\vec{p}_b\times\vec{p}_{\bar b}).
\end{eqnarray}

\item Multi-jet events at the Tevatron: CP-odd correlations. Jets labelled without and with a ``prime'' are associated with the $b$ and $\bar{b}$ jets respectively. Notice that all one needs is to group each $b$ jet with two non-$b$ jets, but it is not necessary to actually distinguish the $b$ jet from the $\bar{b}$ jet.
\begin{eqnarray}
{\cal O}_5 &=& \epsilon(p_b,p_{\bar b},p_{j1},p_{j1'})  \,\,
\xrightarrow[]{b\bar b ~CM}\,\, \propto \,\, \vec{p}_b \cdot (\vec{p}_{j1} \times \vec{p}_{j1'})
\nonumber \\
{\cal O}_6 &=&  \epsilon(p_b,p_{\bar b},p_{j1}+p_{j2},p_{j1'}+p_{j2'})\,\,
\xrightarrow[]{t\bar t ~CM}\,\, \propto \,\, (\vec{p}_{j1}+\vec{p}_{j2})\cdot (\vec{p}_b \times \vec{p}_{\bar b}).
\end{eqnarray}

\end{itemize}

Additional examples can be found in Ref.~\cite{Gupta:2009wu,Gupta:2009eq} or can be easily constructed. The numerical results in these two references also show the relative sensitivity of the many observables to the CP violating couplings.

\section{Numerics}

We used MadGraph \cite{madgraph} to generate all signal and background events both for LHC and Tevatron processes. The signal is calculated separately (analytically) and `hacked' into the MadGraph code. We performed several checks  to satisfy ourselves that this procedure was working correctly. However, for a detector level simulation it would be desirable to be able to generate signal events directly from MadGraph and we are working on this.

\begin{figure}
\vspace{0.2in}
\hspace{1.0in}\includegraphics[width=3.0in]{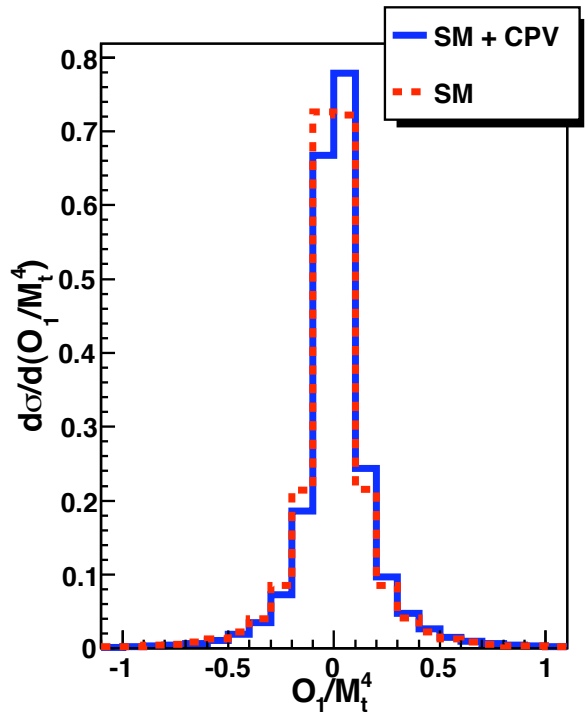}     
\caption{Differential cross-section as a function of the T-odd correlation ${\cal O}_1$. Terms linear in this correlation generate an asymmetry that signals CP violation.}
\label{f:asym}
\end{figure}

An important feature is that there are no background issues for these CP studies beyond those already present in the selection of top-quark pair events. This is because all known backgrounds  are CP conserving. Residual background after event selection will dilute the statistical sensitivity of the signals but will not fake them. It is important to carry out further detector level simulations  to identify potential sources of systematic error. Details of our numerical simulations can be found in the original papers, here we summarize the best results for LHC and put them in perspective in Table~\ref{tab:limits}. An example of one of the distributions is shown in Figure~\ref{f:asym}.

\begin{table}
  \caption{Sensitivity limits at LHC compared to sample models.}
  \label{tab:limits}
  \begin{tabular}{|l|l|l|}
    \hline
    	 coupling & $\tilde{d}\left[\frac{1}{m_t}\right]$ & $\tilde{f}\left[\frac{1}{m_t}\right]$ \\
	 \hline
      Theory estimate & $< 10^{-13}$ SM \cite{review} & 0.03 QCD  \cite{qcdcorr}   \\
      & $\sim 10^{-6} $ with $H^\pm$ \cite{review} & (CP conserving, no phase) \\
      & $\sim 10^{-3}$ SUSY \cite{review} & \\
        \hline
$5\sigma$ sensitivity with 10 fb$^{-1}$ & 0.05     &0.10 \\
\hline
  \end{tabular}
\end{table}

The QCD estimate is for the magnitude of $f$, without any phases. At this level the coupling cannot produce T-odd correlations but if we assume that there are large unitarity phases this number is a rough estimate for the level of the CP even T-odd correlations that appear in the SM. Absorptive phases arise at one-loop in QCD in processes with an additional gluon, and has been considered in  detail in \cite{Hagiwara:2007sz}.

We encourage the experimental collaborations to carry out these measurements both at the Tevatron and at the LHC.

\acknowledgments
This talk is based on work done with J.~F.~Donoghue, Yili~Wang, Oleg~Antipin, Sudhir Gupta, Serhan~Mete and Sehwook~Lee. The original work was supported in part by DOE under contract number DE-FG02-01ER41155. I wish to thank the organizers for a very stimulating workshop, and in particular Fabio Maltoni who also  provided much guidance with the use of MadGraph.


\begin{thebibliography}{99}

\bibitem{tprods}
  J.~F.~Donoghue and G.~Valencia,
  Phys.\ Rev.\ Lett.\  {\bf 58}, 451 (1987)
  [Erratum-ibid.\  {\bf 60}, 243 (1988)]; 
  M.~B.~Gavela, F.~Iddir, A.~Le Yaouanc, L.~Oliver, O.~Pene and J.~C.~Raynal,
  Phys.\ Rev.\  D {\bf 39}, 1870 (1989); 
  M.~P.~Kamionkowski,
  Phys.\ Rev.\  D {\bf 41}, 1672 (1990).
  
\bibitem{ttpairs}  
  J.~P.~Ma and A.~Brandenburg,
  Z.\ Phys.\  C {\bf 56}, 97 (1992); 
  W.~Bernreuther, O.~Nachtmann, P.~Overmann and T.~Schroder,
  Nucl.\ Phys.\  B {\bf 388}, 53 (1992)
  [Erratum-ibid.\  B {\bf 406}, 516 (1993)]; 
  A.~Brandenburg and J.~P.~Ma,
  Phys.\ Lett.\  B {\bf 298}, 211 (1993);
  D.~Atwood, A.~Aeppli and A.~Soni,
  Phys.\ Rev.\ Lett.\  {\bf 69}, 2754 (1992); 
  W.~Bernreuther and A.~Brandenburg,
  Phys.\ Rev.\ D {\bf 49}, 4481 (1994)
  [arXiv:hep-ph/9312210];
  S.~Y.~Choi, C.~S.~Kim and J.~Lee,
  Phys.\ Lett.\  B {\bf 415}, 67 (1997)
  [arXiv:hep-ph/9706379];
  H.~Y.~Zhou,
  Phys.\ Rev.\  D {\bf 58}, 114002 (1998)
  [arXiv:hep-ph/9805358]; 
  J.~A.~Aguilar-Saavedra, J.~Carvalho, N.~Castro, A.~Onofre and F.~Veloso,
  Eur.\ Phys.\ J.\  C {\bf 53}, 689 (2008)
  [arXiv:0705.3041 [hep-ph]];
  Z.~Hioki and K.~Ohkuma,
  arXiv:0910.3049 [hep-ph];

\bibitem{Weinberg:1990me}
  S.~Weinberg,
  Phys.\ Rev.\ D {\bf 42}, 860 (1990).
  
  \bibitem{other} 
  D.~Chang and W.~Y.~Keung,
  Phys.\ Lett.\ B {\bf 305}, 261 (1993)
  [arXiv:hep-ph/9301265]; 
  W.~Bernreuther, A.~Brandenburg and M.~Flesch,
  Phys.\ Rev.\ D {\bf 56}, 90 (1997)
  [arXiv:hep-ph/9701347]; 
  W.~Bernreuther, A.~Brandenburg and M.~Flesch,
  arXiv:hep-ph/9812387.
 

\bibitem{Valencia:2005cx}
  G.~Valencia and Y.~Wang,
  Phys.\ Rev.\  D {\bf 73}, 053009 (2006)
  [arXiv:hep-ph/0512127].

\bibitem{review}
  D.~Atwood, S.~Bar-Shalom, G.~Eilam and A.~Soni,
  Phys.\ Rept.\  {\bf 347}, 1 (2001)
  [arXiv:hep-ph/0006032].

\bibitem{Antipin:2008zx}
  O.~Antipin and G.~Valencia,
  Phys.\ Rev.\  D {\bf 79}, 013013 (2009)
  [arXiv:0807.1295 [hep-ph]].
  
\bibitem{Gupta:2009wu}
  S.~K.~Gupta, A.~S.~Mete and G.~Valencia,
  Phys.\ Rev.\  D {\bf 80}, 034013 (2009)
  [arXiv:0905.1074 [hep-ph]].

\bibitem{Gupta:2009eq}
  S.~K.~Gupta and G.~Valencia,
  Phys.\ Rev.\  D {\bf 81}, 034013 (2010)
  [arXiv:0912.0707 [hep-ph]].
  
\bibitem{delAguila:2002nf}
  F.~del Aguila and J.~A.~Aguilar-Saavedra,
  Phys.\ Rev.\  D {\bf 67}, 014009 (2003)
  [arXiv:hep-ph/0208171].

\bibitem{Sjolin:2003ah}
  J.~Sjolin,
  J.\ Phys.\ G {\bf 29}, 543 (2003);
  
\bibitem{madgraph}
  T.~Stelzer and W.~F.~Long,
  Comput.\ Phys.\ Commun.\  {\bf 81}, 357 (1994)
  [arXiv:hep-ph/9401258]; 
  J.~Alwall {\it et al.},
  JHEP {\bf 0709}, 028 (2007)
  [arXiv:0706.2334 [hep-ph]];
  J.~Alwall, P.~Artoisenet, S.~de Visscher, C.~Duhr, R.~Frederix, M.~Herquet and O.~Mattelaer,
  AIP Conf.\ Proc.\  {\bf 1078}, 84 (2009)
  [arXiv:0809.2410 [hep-ph]].
   
\bibitem{qcdcorr}  
  C.~S.~Li, R.~J.~Oakes and T.~C.~Yuan,
  Phys.\ Rev.\  D {\bf 43}, 3759 (1991).
  
\bibitem{Hagiwara:2007sz}
  K.~Hagiwara, K.~Mawatari and H.~Yokoya,
  JHEP {\bf 0712}, 041 (2007)
  [arXiv:0707.3194 [hep-ph]].

 
\end{thebibliography}
\end{document}